\documentclass{article}

\usepackage[english]{babel}

\usepackage[letterpaper,top=3cm,bottom=3cm,left=3cm,right=3cm,marginparwidth=1.75cm]{geometry}
 \usepackage{setspace}
\usepackage{array}
\usepackage{booktabs}
\usepackage{multicol}
\usepackage{amsmath}
\usepackage{graphicx}
\usepackage{natbib}
\usepackage{tabularx}
\usepackage[dvipsnames]{xcolor}
\usepackage[colorlinks=true, allcolors=blue]{hyperref}

\newcolumntype{R}[1]{>{\raggedright\arraybackslash}p{#1}}
\title{A Computational Social Science Approach to Understanding Predictors of Chafee Service Receipt}
\author{Jason Yan,\textsuperscript{a} Melanie Sage,\textsuperscript{b} Yuhao Du,\textsuperscript{a} Seventy F. Hall,\textsuperscript{b} Kenneth Joseph\textsuperscript{a,1}\\ {\small \textsuperscript{a} Department of Computer Science and Engineering, University at Buffalo} \\ 
{\small \textsuperscript{b} School of Social Work, University at Buffalo} \\
{\small \textsuperscript{1} Corresponding Author: kjoseph@buffalo.edu}
}
\date{}

\begin{document}
\maketitle

\begin{abstract}
The John H. Chafee Foster Care Program for Successful Transition to Adulthood (CFCIP) allocates funding to provide services to youth who are likely to age out of foster care. These services, covering everything from mentoring to financial aid, are expected to be distributed in ways that prepare youth for life after care. However, surprisingly little is known about which youth receive which services.  The present work makes use of the National Youth in Transition Database (NYTD), a large-scale administrative dataset that tracks services allocated to youth that use CFCIP funds. Specifically, we conduct a \emph{forensic social science} analysis of the NYTD data. To do so, we first  use computational methods to help us uncover the most important factors associated with service receipt.  Doing so helps us to identify three major factors---youth age, youth time in care, and the state in which a youth is in care---that are most heavily associated with service receive. We then conduct an analysis that links existing theory to these factors, expanding our understanding of how services are allocated and paving the way to future work to understand \emph{why} such associations exist.
\end{abstract}

\section{Introduction}

The primary stated goal of the U.S. child welfare system (CWS) is to keep youth with their families. Despite this goal and efforts to achieve it, however, tens of thousands of youth each year are removed from their families. Many of these youth will eventually \emph{age out} of foster care without returning to their families or being adopted; that is, they will turn an age at which they are no longer considered wards of the state and eligible for services. Youth who age out of care, on average, have significantly worse life outcomes as compared to general populations of youth. For example, they are known to experience criminal or juvenile justice system involvement, food insecurity, and homelessness at much higher rates than their peers \citep{lockwood2015permanency}.

Recognizing this, the federal government has passed several laws that aim to improve outcomes among older youth in foster care. The Foster Care Independence Act, signed into law in 1999, established the John H. Chafee Foster Care Program for Successful Transition to Adulthood (CFCIP). Under CFCIP, the federal government allocates Title IV-E funding to states to provide services to foster youth who are likely to age out of care. States in turn use this money to refund foster care agencies, both public and private, for services rendered to foster youth. Services funded by the Chafee program include anything from mentoring to financial assistance for housing and post-secondary education.

A number of studies have explored how services are allocated to foster youth under the CFCIP \citep{chorYouthSubgroupsWho2018,perezFactorsPredictingPatterns2020,okpychReceiptIndependentLiving2015,thompsonFactorsPredictingService2021}.
 Doing so is important in order to a) identify potential biases in how services are allocated, and b) develop prescriptive recommendations for better or more refined intervention in the future. One important resource of more recent research in this area is the National Youth in Transition Database (NYTD) \citep{nytdcite}.  Foster care agencies that receive funding from CFCIP are mandated to report data regarding case level information on the youth and the independent living services they received. These services are entered into the NYTD, coded into one of fifteen different service types that the database tracks. Data on youth service receipt are collected and recorded bi-annually in the NYTD, giving a longitudinal snapshot of allocation patterns over time.

The first and most widely cited use of the NYTD data to study service allocation patterns comes from  \citet{okpychReceiptIndependentLiving2015}. They analyzed Chafee service receipt across the U.S. among all youth ages 16-21 who were in foster care in fiscal year (FY) 2011 and who were eligible for Chafee-funded services. Their goal was to identify characteristics of youth who were eligible for the Chafee program who in turn did, or did not, actually receive services.  Using descriptive statistics, their work shows that receipt of specific services varies across race/ethnicity, gender, and whether or not an individual has been identified as having a disability. Their work also shows important effects on the urbanicity of the county in which a youth is in care, and that this effect of urbanicity varies across racial lines. 

The present work provides a complementary analysis of NYTD data on Chafee service receipt to the work of \citet{okpychReceiptIndependentLiving2015}. The core difference between our work and theirs is in the analytical strategy employed. Specifically, our work draws on an analytical paradigm increasingly popular in sociology \citep[e.g.][]{salganikMeasuringPredictabilityLife2020,kozlowskiGeometryCultureAnalyzing2018}, social psychology \citep[e.g.][]{gargWordEmbeddingsQuantify2018}, and political science \citep[e.g.][]{robertsStructuralTopicModels2014} that is commonly referred to as \emph{forensic social science} \citep{mcfarlandSociologyEraBig2015}.  Forensic social science ``is an approach that merges applied and theory-driven perspectives...[where we] use theory to partly guide deductive explorations of the data while also using induction to discover which theories afford an explanation" \citep[][pg. 20]{mcfarlandSociologyEraBig2015}.  Critically, forensic social science, as a subfield of computational social science \citep{lazerComputationalSocialScience2009,epsteinAgentbasedComputationalModels1999}, is distinct from forensic \emph{social work}, an entirely different methodology defined as "the practice specialty in social work that focuses on the law and educating law professionals about social welfare issues and social workers about the legal aspects of their objectives"  \citep[][p. 140]{barker2003social}, \citep[cf.][]{robertsCenturyForensicSocial1999}. 
Where a more traditional social scientific approach, like that pursued by \citet{okpychReceiptIndependentLiving2015}, uses a small number of independent variables to evaluate a finite set of hypotheses, forensic social science instead aims to leverage "big data" and modern computational methods to confirm existing claims in new ways, and to \emph{generate} new avenues for theory that were not apparent before the analyses were conducted \citep{goldbergDefenseForensicSocial2015}. 


We employ a particular kind of forensic social science analysis here that uses predictive modeling to better understand social phenomenon \citep[see][discussed below, for the most elaborate example]{salganikMeasuringPredictabilityLife2020}. Notably, the use of predictive modeling within the paradigm of forensic social science is distinct from its use in what we will call \emph{prescriptive predictive modeling}, where the goal is to use predictive modeling to \emph{make or aid in on-the-ground decisions}. While forensic social science may advance our ability to make decisions, it does so by providing evidence for or new pathways to social theory and future research that can more directly pinpoint causal mechanisms and thus identify potential policy changes. 

With a predictive modeling approach, we first identify a \emph{prediction task}. To do so, we first identify a population of interest, and an outcome variable that we wish to study. Our prediction task then becomes the task of predicting the value of this outcome, given other information available to us about our population.  Having identified a prediction task, we then develop and carry out a \emph{prediction experiment}, in which we evaluate a number of different approaches to predicting the outcome variable. These approaches, as we will see, range in both the complexity and quality of their predictions. Our experiment consists of repeated trials, where we separate our population into "training" and "test" sets, learning parameters of computational models (e.g. fitting a regression) on the training data, and then estimating prediction quality ("model performance") on the test data.  Finally, we engage in \emph{model exploration}, where we use existing theory, combined with the results of the prediction experiment and the parameters of the estimated models, to understand a) the relative importance of various factors in predicting our outcome, and b) the limits of how well we can make such predictions, given the data we have.

In the present work, our prediction task is as follows: \emph{given data on foster youth from the NYTD and other associated datasets, can we predict how many distinct services that youth will receive in a given year?} Informed by prior work and empirical observations, we ask this question across three distinct subsets of services.  Our predictive experiment then compares a number of models ranging from simple baselines (e.g. predicting the same number for all youth), to standard  regression models, to \emph{machine learning} models that incorporate dozens of predictors into a complex, non-linear function to make predictions. Finally, our model exploration evaluates factors that appear critical across these various models, the relative predictive power of various individual-level and structural factors, and important limits on how well services can be predicted from administrative data.

In doing so, our work makes the following three contributions to the literature:
\begin{itemize}
    \item In the context of Chafee service allocation, we show that independent of other factors, \emph{age}, \emph{length of stay}, and \emph{state of care} are the strongest predictors of the number of services youth receive. We provide a discussion of the implications of these findings in the context of existing knowledge of how services are allocated.
    \item In the context of analytical methods in child welfare, we further establish forensic social science as a principled methodology that, when used in complement with theory-driven scholarship, can provide new insights into fundamental problems \citep{rodriguezComputationalSocialScience2020}.
    \item At the same time, our work reinforces prior work emphasizing caution in using predictive modeling as a tool for direct decision-making \citep{chouldechova2018case}, rather than as an integrated element of theory development. We do so by highlighting how more complex models can, if used to make allocation decisions, further existing inequalities in child welfare. 
\end{itemize}

\section{Background}

\subsection{Prior work on Chafee Service Allocation}

A number of recent efforts have been made to study how Chafee services are allocated using the NYTD. As mentioned previously, \citet{okpychReceiptIndependentLiving2015} analyzes Chafee service receipt across the U.S. among youth in foster care (ages 16–21) using data from the NYTD from 2011 and 2012. Findings included observations of biases in service allocation across gender---females were more likely to receive services, race---multiracial and Hispanic youth were more likely to receive at least one service than any other groups, and Black youth were least likely to receive a Chafee service--- and county urbanicity--youth in rural/nonmetropolitan areas are more likely to receive more services, including more kinds of services, than youth in large metropolitan areas.

Recent work has moved beyond the descriptive efforts of \citet{okpychReceiptIndependentLiving2015}. In particular, \citet{chorYouthSubgroupsWho2018} study youth in foster care from FY 2011-2013 who received at least one Chafee-funded service according to the NYTD. The authors use a variant of latent class analysis (LCA), multi-level LCA, that accounts for state-level effects to cluster youth based on the set of services they received. Their work identifies three classes of service profiles for youth in their data:  a "High-service receipt" class, a "Limited service receipt" class, and a class of youth who only received Academic Support or an Independent Living Needs assessment. These classes varied in size, representing around 20\%, 30\%, and 50\% of their data, respectively. \citet{chorYouthSubgroupsWho2018} then show significant predictors of youth falling into each class, finding differences on age, gender, educational attainment, and race. \citet{perezFactorsPredictingPatterns2020}, using the same methodology as \citet{chorYouthSubgroupsWho2018}, but with only 16-year old youth, identify a similar three-class clustering of youth by service receipt, and similar underlying factors predicting which of these clusters youth fall into.


The present work compliments these prior efforts in several ways. With respect to \citet{okpychReceiptIndependentLiving2015}, we use a similar sampling scheme to define our sample population for study, but extend their findings to a number of other independent variables that are associated with service allocation.  And, while we consider the efforts of \citet{chorYouthSubgroupsWho2018} and \citet{perezFactorsPredictingPatterns2020} to be similar in the use of computational methods to surface observations for future hypothesis-directed work, our work extends their efforts by 1) including into our analysis youth who did not receive any services, in addition to those that did, 2) significantly extending the set of independent variables considered, and 3) using more recent iterations of the NYTD dataset.

\subsection{Forensic Social Science}

The utility of a forensic social science approach is perhaps best described by \citet{mcfarlandSociologyEraBig2015}:
\begin{quote}
[T]he use of machine learning is atheoretical, but it is potentially powerful when used as an agnostic search for potential explanations. In contrast, theory is a somewhat narrow-minded but powerful tool... it is a focusing device that identifies which constructs are to be selected and formed from the millions of possible variables (or features) and it afford potential explanations for how features interrelate. As such, the iterative combination of atheoretical induction and theory-led deduction can be quite powerful. (pg. 10)
\end{quote}

Perhaps the most well known empirical forensic social science work is from \citet{salganikMeasuringPredictabilityLife2020}. In this work, the authors organize and conduct a competition to predict life outcomes for youth in the Fragile Families dataset, which tracks youth from birth through adulthood on a variety of life outcomes.  The core finding of the challenge was that despite dozens of complex models used in the competition, nothing systematically outperformed a simple,  baseline model that used a simple regression to predict life outcomes at age 18 from four theoretically-motivated variables: mother's race/ethnicity, marital status, and education level, and the same outcome (or a closely related proxy) measured at age 9.  This finding informs our work in two ways. First, we ensure that we compare our predictive model results to similar, simple baseline models in order to determine whether or not more complex models might help in better understanding service allocation. Second, as we find that more complex models \emph{do} perform better than our baseline models, we detail potential reasons why service allocation may be a setting in which predictive modeling is particularly useful for understanding decision-making patterns.

In addition to this empirical work, scholars have begun to more concretely formalize ways in which forensic social science can formally integrate theory and machine learning together in a single analysis.  \citet{nelsonComputationalGroundedTheory2020}, for example, proposes such a framework for the analysis of text data, and \citet{rodriguezComputationalSocialScience2020} extends this approach and shows various markers of validity. \citet{radfordTheoryTheoryOut2020a} present a more general case for how social theory and machine learning can, and must, be used together to inform forensic social science analyses. The present work is informed by these efforts, in that we 1) are careful to align findings from our predictive experiments with prior theory and research in social work, and 2) explore in our discussion ways that our findings present new possibilities for theory and further analysis, rather than being of causal importance on their own.

\subsection{Predictive Modeling in Child Welfare}

As noted above, forensic social science uses predictive models to help inform  how services are allocated. This differs significantly from the standard use of predictive analytics in child welfare today, where scholars aim to build algorithms that can be used in the decision-making process.  For a general review of this usage, we direct the reader to \citet{saxenaHumanCenteredReviewAlgorithms2020}, here we provide a more brief overview.

Existing efforts to use algorithms in child welfare have primarily attempted to predict the level of risk when a child is referred to a child welfare hotline. Predictions are typically based on historical administrative data from public welfare systems \cite{teixeira_predictive_2017}. This use of algorithmic decision making has elicited significant critiques given the possibility that past data embeds the bias of past assessments \cite{capatosto_foretelling_2017}, and is impacted by the over-surveillance of families of color and those who use public resources (e.g. public welfare) \cite{eubanks_automating_2018,glaberson_coding_2019}. 

In child welfare systems and other public benefit bureaucracies, the risks and rewards of using algorithms are particularly impactful to service users, who are often already the most vulnerable people in our society, and therefore these algorithmic methods deserve additional scrutiny.  Human services administrators often have little training in predictive analytics, and may turn over decision-making to vendors who make lofty promises of efficiency and cost-savings; vendors may aksi deliver black-box algorithms that under-perform \cite{kelly_illinois_2017}. These algorithms often fail to fully account for the cultural context in which the data is collected and the ways that algorithmic decisions are made \cite{church_search_2017} and/or used \cite{christin_algorithms_2017}. Significant concerns thus arise with algorithms that, while meeting algorithmic definitions of fairness, do not necessarily meet practical definitions of fairness that control for issues such as equity, transparency, or the degree with which non-white youth are over-represented in the child welfare system and suffer disproportionately poorer outcomes \cite{brown_toward_2019}. 

These findings from prior work point to a need for extreme caution in using predictive analytics in the context of child welfare. Here, we are motivated by the potential for predictive analytics to serve not in a decision-making role, but rather, as with prior work using statistical models on administrative data reviewed above, to better understand and, in turn, theorize, the ways that children are impacted by decision-making in existing system.

\section{Data}

Our analysis draws on data made available by the National Data Archive on Child Abuse and Neglect (NDACAN).  Here, we leverage two specific datasets made available by NDACAN.  The first is data derived from the Adoption and Foster Care Analysis and Reporting System(AFCARS) \citep{afcarscite}. AFCARS provides, on a yearly basis, information about all youth that have been in foster care, i.e. any youth that meets the federal definition of 45 CFR 1355.20 in that particular year.  Second we use data from the NYTD. The NYTD is comprised of multiple datasets; here we use only the data that contains a public record of all youth who received Chafee-funded services that were reported to the state. This NYTD \emph{Services data} is released twice per year, and can be linked through an anonymized ID to the AFCARS dataset. Although they represent the most complete datasets related to demographics and service allocation for youth in foster care, it has limitations related to data accuracy, which youth are represented in the NYTD dataset, and how service-related variables are defined.

Our analysis is focused primarily on data from the 2018 fiscal year (FY).
In this section, we first describe how we select individuals from the full AFCARS dataset for inclusion in our study.  We then describe three dependent variables that will serve as the outcomes we seek to predict. Finally, we detail the various independent variables (or synonymously here, \emph{features}) that we will use to make these predictions.

\subsection{Inclusion/Exclusion Criteria}

\begin{table}[]
    \centering
    \begin{tabular}{|p{9cm}|>{\raggedleft\arraybackslash}p{2.5cm}| >{\raggedleft\arraybackslash}p{2.5cm}|}
    \toprule
 \textbf{Inclusion Criterion}  & \textbf{N Removed} & \textbf{\%  Removed}  \\ \toprule
   Younger than 22 at the start of the FY   & 30 & $<$.001\%  \\
   Older than 14 at the start of the FY & 552,641 & 80.4\% \\
   In care for at least 6 months in FY & 49,189 & 36.5\% \\
   First removal date $>$22 years prior to FY & 13 & .02\%  \\
   Removal or Setting Change Dates Not Missing & 85 & .1\%  \\
   Race/Ethnicity Data Not Missing & 883 & 1.0\% \\
   In State where Services were Recorded (not NC or PR) & 2,172 & 2.6\% \\
   Setting Length of Stay Not Missing & 931 & 1.1\% \\
   Parents Died Indicator Not Missing & 710 & 0.9\% \\ \midrule
   \textbf{Total youth removed from study} & 606,654 & 88.3\% \\ \midrule
   \multicolumn{3}{c}{{\bf Total Youth Included: 80,714}} \\ \midrule
    \end{tabular}
    \caption{A sequential list of inclusion criterion used in this study to identify the final population of interest. The first column in this table lists the criteria used. The second lists how many youth were removed from the sample because of that criteria. The final lists the percentage of the remaining sample removed because of this criteria (i.e. the percent of the sample that resulted from all previous criteria). }
    \label{tab:inclusion}
\end{table}

Table~\ref{tab:inclusion} lists the set of inclusion criterion that youth were required to meet to be considered for our study. The primary cause of exclusion from the study was age; in particular, following prior work we analyze only youth aged 14-22. These youth received the vast majority (over 98\%) of all services. The second most important exclusion criterion was ensuring that youth were in care for at least 6 months; this was to ensure that we analyzed youth who had spent enough time in care to have had the opportunity to receive services if they were likely to get them. Finally, we excluded youth based on a number of missing data criterion, where these missing data did not seem reasonable to impute (see below), and excluded youth in North Carolina and Puerto Rico, where Chafee services data were not recorded in FY 2018 in the NYTD dataset.

\subsection{Dependent Variables}

Our prediction experiment explores three different outcome variables drawn from fourteen of the fifteen services listed in the NYTD Service File.  Excluded from our analysis entirely are Special Education services. We exclude these for two reasons. First, special education services are \emph{theoretically} distinct from the other services because they are associated with a school-based assessment instead a social services assessment.   Second, as we show below, special education is \emph{empirically} distinct from the other services allocated. With the remaining fourteen services, we construct three different dependent variables:
\begin{enumerate}
    \item \textbf{All Services} - The total number of unique services that a youth receives. The maximum possible value for this outcome is 14; this includes all services in the NYTD dataset, except for Special Education (these are listed on the x-axis of Figure~\ref{fig:services}).
    \item \textbf{Financial Services} - The number of unique services a youth receives from the following set: Supervised Independent Living, Room and Board Financial Assistance, Education Financial Assistance, Other Financial Assistance. These services are unique in that they all either pay a youth directly in cash or pay for a service that the youth would normally pay for themselves to meet everyday needs related to housing and education.
    \item \textbf{Academic and Employment Support Services} The number of unique services a youth receives from the following set: Academic Support, Post-secondary Educational Support, Career Preparation, Employment Programs or Vocational Training, Budget and Financial Management, Housing Education and Home, Health Education and Risk Prevention, Family Support and Healthy Marriage Education, and Mentoring. These services all provide non-monetary supports, generally in the form of education or social support.
\end{enumerate}

These three sets of outcome variables are theoretically interesting for different reasons. All services, together, help us understand factors related to who gets the most and least services. Financial services help us understand services provided directly to youth to practice their own independence, and are often offered to youth who are expected to pay for their own needs, whereas academic and employment support services  often  seek to fill knowledge gaps through training programs to promote self-sufficiency, and are sometimes corrective. For instance, family support and education services are often delivered to youth who are already parents or are likely to become parents.  In addition to the theoretical reasons for selecting these subsets of services, we again show below that there is also empirical support that these services are clustered together in their allocation as well.

\subsection{Independent Variables}

\begin{table}[t]
\scriptsize
    \centering
    \begin{tabular}{p{4cm}|p{11cm}} 
    \toprule
\multicolumn{2}{c}{\textbf{Categorical Variables}} \\ \toprule 
\textbf{Variable Name} & \textbf{Variable Levels} \\ \midrule
Race/Ethnicity & NH, White; NH, Am Ind AK Native; NH, Asian; NH, Haw/Pac Isl.; NH, $>$1 Race; Hispanic \\
Case Goal &  Reunification; Live w/ Relatives; Adoption; Long Term Foster Care; Emancipation; Guardianship; Not Yet Established \\  
Caretaker Family & Married Couple; Unmarried Couple; Single Female; Single Male; Unknown/None \\
Placement Setting & Pre-adoptive Home; Foster family home,relative; Foster family home, non-relative; Group home; Institution; Supervised independent living; Runaway; Trial home visit \\
Manner of Removal & Removed Voluntarily; Court Ordered; Unknown Reason for Removal \\
RU13 Urban/Rural Code & 8 Levels, Ranging from 1) Metro: $>$1 million population to 8) Rural or $<$ 2.5K population, Non-Adjacent \\ 
State & All 50 states, besides North Carolina, Plus DC \\
Age (Categorical) & 14, 15, 16,17, 18, 19+ \\
Age on Date of Legal Adoption & Not Applicable; Less than 2 years old; 2-5 years old; 6-12 years old; 13 years or older; Unable to determine \\
Discharge Reason & Not Applicable; Reunified with parent, primary caretaker; Living with other relative(s); Adoption; Emancipation; Guardianship; Transfer to another agency; Runaway; Death of child \\
Sex & Male, Female \\
\toprule
\multicolumn{2}{c}{\textbf{Binary Variables} (3 Levels for each: Yes/Applies; No/Does not Apply; Unknown)} \\ \toprule
\multicolumn{2}{p{15cm}}{Sexual Abuse; Physical Abuse; Neglect; Alcohol Abuse Parent; Drug Abuse Parent; Alcohol Abuse Child; Drug Abuse Child; Child Disability; Child Behavioral Problem; Parents Died; Parents in Jail; Caretaker inability; Abandonment; Relinquishment; Inadequate Housing; Parents Rights Terminated; Title IV-E Foster Care Payments; Title IV-E Adoption Assistance; Title IV-A TANF Payment; Title IV-D Child Support Funds; Title XIX Medicaid; SSI or Social Security Act Benefits; Only State or Other Support; Aged Out of Foster Care; Clinical Disability, Mental Retardation; Visual/Hearing Impaired; Physical Disability; Emotionally Disturbed; Other Medical Issue; Current Placement Setting Outside State; Dad's Rights Terminated; Mom's Rights Terminated, In AFCARS Dataset in Previous Year; Was Discharged From Latest Removal; Was Discharged From Previous Removal; Child is Waiting for Adoption; Has Ever Had Periodic Review}   \\
 \toprule
\multicolumn{2}{c}{\textbf{
Continuous Variables} (Both Raw Values and Logged Values)} \\ \toprule
\multicolumn{2}{p{15cm}}{Previous year service count (all services); Previous year service count (Academic and Employment Support Services); Previous year service count (financial services); Previous year service count (each service individually); Current Setting Length of Stay; Total Number of Removals; Total Number of Placements; First Removal Date; Latest Removal Date; Latest Setting Date; Date of Discharge from Previous Removal; Date of Discharge from Latest Removal; Date of Latest Periodic Review; Age at End of FY} \\
    \end{tabular}
    \caption{A Summary of all variables used in our models. For this table, we categorized variables into three types: Categorical Variables with meaningful variable levels, binary variables (with an "Unknown" level in some cases), and continuous variables. These three types of variables are shown separately. Categorical variables are listed alongside their various levels; binary and continuous variables are simply listed. Variables and variable levels are separated by semicolons (;).}
    \label{tab:variables}
\end{table}

Table~\ref{tab:variables} lists all 61 categorical, binary, and continuous variables used in our analysis to attempt to predict our dependent variables.  Descriptive statistics for each variable are included in the supplementary material provided in the code release for this project. For continuous variables, we include both the raw value and the logarithm of the variable in our predictive models, as in common in predictive analyses where both exact time frames and orders of magnitude for time may provide salient information \citep{salganikMeasuringPredictabilityLife2020}. All continuous variables relevant to dates are measured as days since the end of the fiscal year.

Missing values are addressed in one of two ways. Where a reasonable default value could be identified, we replaced missing values with appropriate defaults. For example, 2,380 youth had missing values for Clinical Disability; these rows were replaced with a value representing "Not Yet Determined". The full set of imputed variables are provided in the data and code accompanying this article.  In a related vein, we do not attempt to impute values for categories marked as "Unknown" or "Not Yet Determined". The primary reason for this is that we do not believe these values to be missing at random, but rather that the missingness of these values is potentially a meaningful signal of how a youth is receiving services within the foster care system. Rather than seek to infer "true" values for these missing quantities, then, we treat missingness itself as a theoretically and empirically meaningful quantity.

\section{Methods}

We now move to a description of our prediction experiment and our approach to model exploration.  We first describe the set of \emph{predictive models} that we construct to make predictions about the number of services youth will receive. Note that in all cases for these predictive models, model parameters are calculated on one subset of the data, the \emph{training data}, and evaluated on another (the \emph{test data}).  We then detail how we \emph{evaluate} these models using evaluation metrics computed over repeated splits of training and test data. Finally, we detail our methodology for model \emph{exploration}.

\subsection{Prediction Models}

We construct two machine learning models to make predictions about service allocation.  Both of our machine learning models are  \emph{tree-based}. We opt to focus on tree-based models for two complementary reasons. First, tree-based models are capable of capturing the complex dependencies inherent in our administrative data.  An example of this is the variable \emph{Date of Latest Periodic Review}, which is dependent on the variable \emph{Has Ever Had Periodic Review}. In a tree-based model, the algorithm can "learn" to separate youth into those that have or have not ever had a periodic review, and then make use of information contained in the date of that review. Notably, a linear model (e.g., a regression), cannot do this, even with an interaction term, because a coefficient for each independent variable is applied to each youth.  

Second, while other machine learning models can handle these forms of dependencies, tree-based models do so in a way that is (relatively more) straightforward to explain. Our first tree-based model is the \emph{Random Forest} \citep{breiman2001random}. A random forest model is an ensemble of decision trees, where the output of the random forest is the aggregate of all decisions made by the individual decision trees. In the case of a random forest regressor, which we used in our work, the output is simply the average of all the individual decision trees estimate. The idea behind a random forest is that individual decision trees may not be so accurate, but when combined together, the output will be closer to the true value on average. Our second tree-based model is the \emph{Gradient Boosted Trees (GBT)} algorithm \citep{friedman2001greedy}, implemented in the computing library \texttt{XGBoost}. It is a gradient boosting algorithm that iteratively combines decision trees as the "weak" predictors to produce a much stronger model. The principal idea is that each decision tree builds upon the previous trees by learning the residuals, essentially a correcting term. The final output is the sum of the output from each individual tree. 

Both the random forest model and the GBT model we use contain \emph{hyperparameters}, or parameters that are fixed for any one run of the algorithm. There are a number of ways to select hyperparameters. Here, we choose to do so in two different ways to avoid dependency on any one strategy. For random forests, we identify hyperparameters by finding the hyperparameters that allow for the best predictions on services data from prior years. We then fix these hyperparameters for prediction on the 2018 data.\footnote{The hyperparameters we selected were: 1000 decision trees with the max depth of any tree being 15 and the minimum number of samples required to split a node is 12. We also require that the minimum number of samples in a leaf node is 5. Bootstrap samples are used to build each individual trees.}. For GBT, rather than optimize hyperparameters on our data, we instead opt to select a single setting for the hyperparameters that prior work has suggested is a reasonable default.\footnote{The hyperparameters we selected were: 0.15 for the learning rate, zero loss reduction required for a node split, 70 percent of the features are randomly selected when building each individual trees. Minimum samples to split a node is 3, and the max depth of any tree is 6.}

In addition to these two models, we construct a number of simple and more complete baseline models in order to evaluate prediction quality. We first detail six simple baseline models derived from reasonable expectations based on prior work described above about 1) how past service receipt should predict future receipt, 2) the impacts of age, race, and rural/urbanicity \citep[as considered by][]{okpychReceiptIndependentLiving2015}, and 3) the impact of the state in which a youth lives on the services they receive \citep[as considered by][]{chorYouthSubgroupsWho2018,perezFactorsPredictingPatterns2020}:
\begin{enumerate}
    \item \emph{Previous Year's Service Count} - For this baseline model, we predict the number of services a youth will receive in 2018 using the number of services they received in 2017. If a youth was not in care in 2017, we predict a value of 0. 
    \item \emph{Constant} - We first compute the average number of services received by youth in the training data. We then predict that each youth in the test data will receive this amount of services. This model is equivalent to a linear regression model with only an intercept term.
    \item \emph{Age} - We first compute the average number of services received by youth in each age category (14, 15, 16, 17, 18, 19+) in the training data. We then use these averages to make predictions in the test data, based on the age category of each youth. This model is equivalent to a linear regression model with a single predictor (age category).
    \item \emph{Age and Race} - We employ the same approach as above, except we compute an average for each combination of age category and race/ethnicity category. This model is equivalent to a linear regression model where each combination of age and race/ethnicity is included as a predictor.
    \item \emph{Age, Race, and RU-13} - We employ the same approach as above, except we compute an average for each combination of age category, race/ethnicity category, and RU-13 county designation. This model represents the most complex descriptive statistics reported by \citet{okpychReceiptIndependentLiving2015}.
    \item{\emph{Age and State}} - We employ the same approach as above, except we compute an average for each combination of age category and state.
\end{enumerate}

As noted, the baseline models used can be interpreted as simple linear regression models, akin to those used as a baseline model by \citet{salganikMeasuringPredictabilityLife2020}.  However, these baseline models are simplified both in the independent variables used, and in the assumption of a linear relationship between these variables and the outcome. To this end, it is also useful to compare our models to baselines that use all of the same independent variables, differing only on the linearity assumption.

To this end, we include as additional baselines two regression models. The first is a linear regression model, which uses all of the same variables as our tree-based models (those identified in Table~\ref{tab:variables}). Because the outcome is a count variable we also use a negative binomial model. In addition, for the negative binomial model we incorporate a set of fixed effects for each age/state combination, as one would traditionally do for a traditional statistical analyses of the data at hand. \footnote{ Notably, another appropriate model is a zero-inflated Poisson model. Attempts to estimate this model using our data were unsuccessful unless we limited the independent variables to a subset of those listed in Table~\ref{tab:variables}; as such, we do not include those results here.}

\subsection{Evaluation}

Our evaluation is conducted using a procedure known as \emph{K-fold cross validation} \citep{mosteller1968data}. To do so, we first split our data up into K subsets, here, $K=10$. We then select the first subset of the data and treat it as the test data, using the remaining (90\%) of the data for training. We then repeat this same process using the second subset of the data as the test data, and then the third, and so on. We therefore train and evaluate the same models ten different times, allowing us to obtain confidence estimates on our evaluation metrics.

We use two statistics to evaluate a given set of predictions from a given predictive model. The first metric is the \emph{Root Mean Squared Error (RMSE)}, the most common measure of error in prediction of continuous quantities. Formally, given a set of $N$ youth, the RMSE is computed as $ \sqrt{ \frac{\sum_i^N (\hat{y_i}-y_i)^2}{N}}$, where $\hat{y}_i$ is the predicted number of services for the \emph{i}th youth in the dataset, and $y_i$ is the true number of services that youth received.  The second metric is designed to evaluate potential inequalities in how services would be allocated if a predictive model were used to actually allocate services. To this end, we compute the \emph{mean error}, $\sum_i^N \frac{\hat{y_i}-y_i}{N}$, for each model, for each race/ethnicity of youth.

\subsection{Model Exploration}

Once establishing the most predictive model, we then explore what this model can tell us about the factors that are associated with your receiving more (or less) services. In a traditional regression model, we can assess which variables are most predictive by simply looking at regression coefficients. In more complex models, however, or in regression models with high levels of collinearity, it becomes a challenge to determine the impact of any one fact on the predictions of a given predictive model.

To address this challenge, scholars have constructed various methodologies; a review of which can be found in \citet{}. Here, we adopt one of the most popular such methods, called \emph{SHapley Additive exPlanation (SHAP) values} \citep{lundberg2017unified}. SHAP values are quantities that can be computed for each independent variable for each youth. The SHAP value for an independent variable for a particular youth represents the expected change in the outcome variable (here, the number of services the youth with receive) given that youth's value for that independent variable.  Aggregated, or analyzed, over all youth, SHAP values can therefore give a sense of the way in which a change in a given independent variable impacts the number of services a youth receives.

\section{Results}

We begin by presenting a number of descriptive statistics that provide further insight into the dependent variables chosen for our analysis.  We then devote sections to describing the results of the predictive experiment we conducted, as well as an exploration of the best-performing model in terms of RMSE.

\subsection{Descriptive Statistics}

\begin{figure}[t]
    \centering
      \includegraphics[width=.7\textwidth]{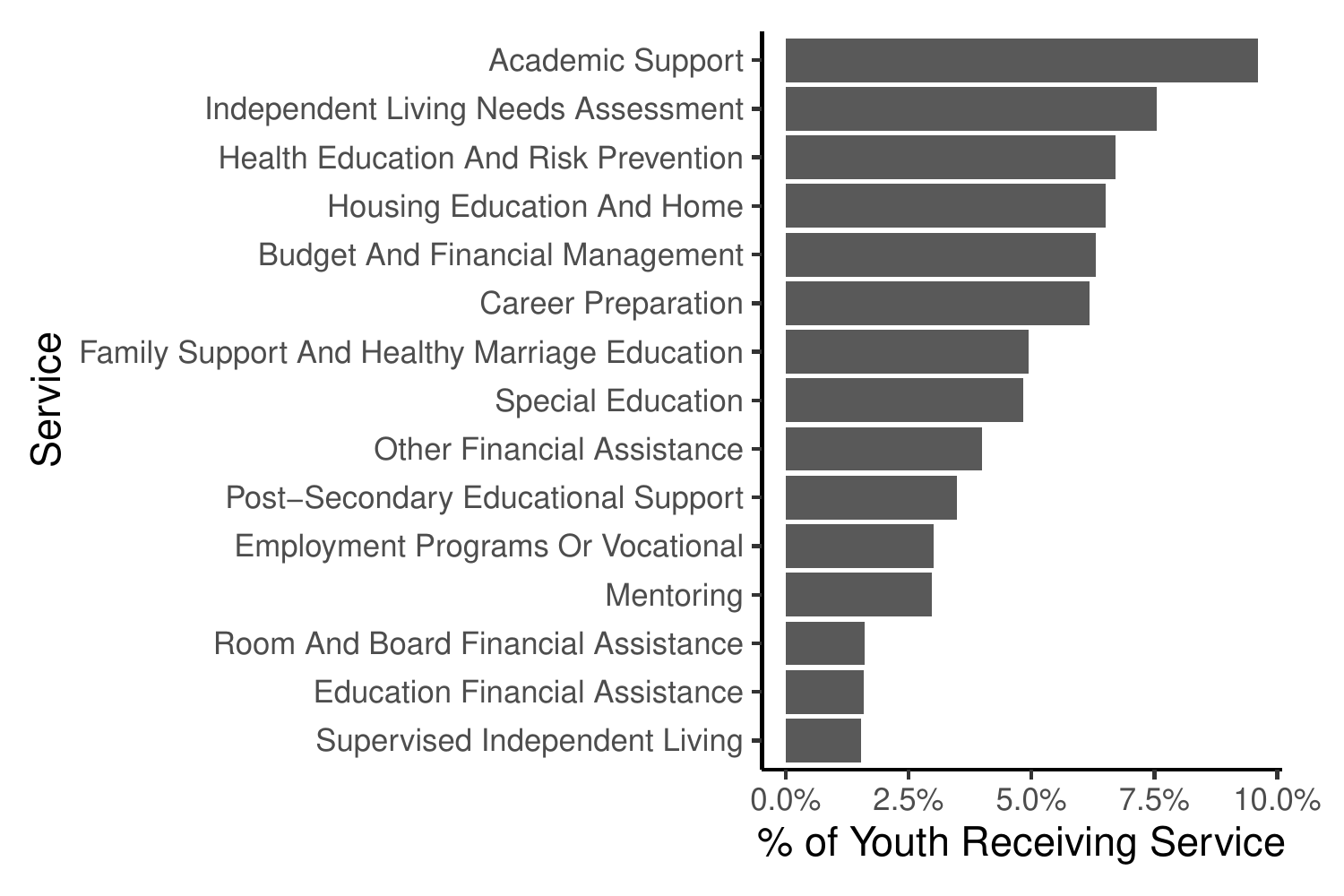} 
    \caption{The percentage of youth (y-axis) listed as having received each Chafee service (y-axis) at least one time in FY 2018.}
    \label{fig:services}
\end{figure}

Figure~\ref{fig:services} lists each of the 15 services in the NYTD data, along with the percentage of youth that received each service. Academic support and independent living needs assessments were the most frequent, with 9.6\% and 7.5\% of youth receiving these services, respectively. The least frequently received services were Supervised Independent Living, Room and Board Financial Assistance, and Educational Financial assistance. Only around 1.5\% of youth received these services.

\begin{figure}[t]
    \centering
      \includegraphics[width=.8\textwidth]{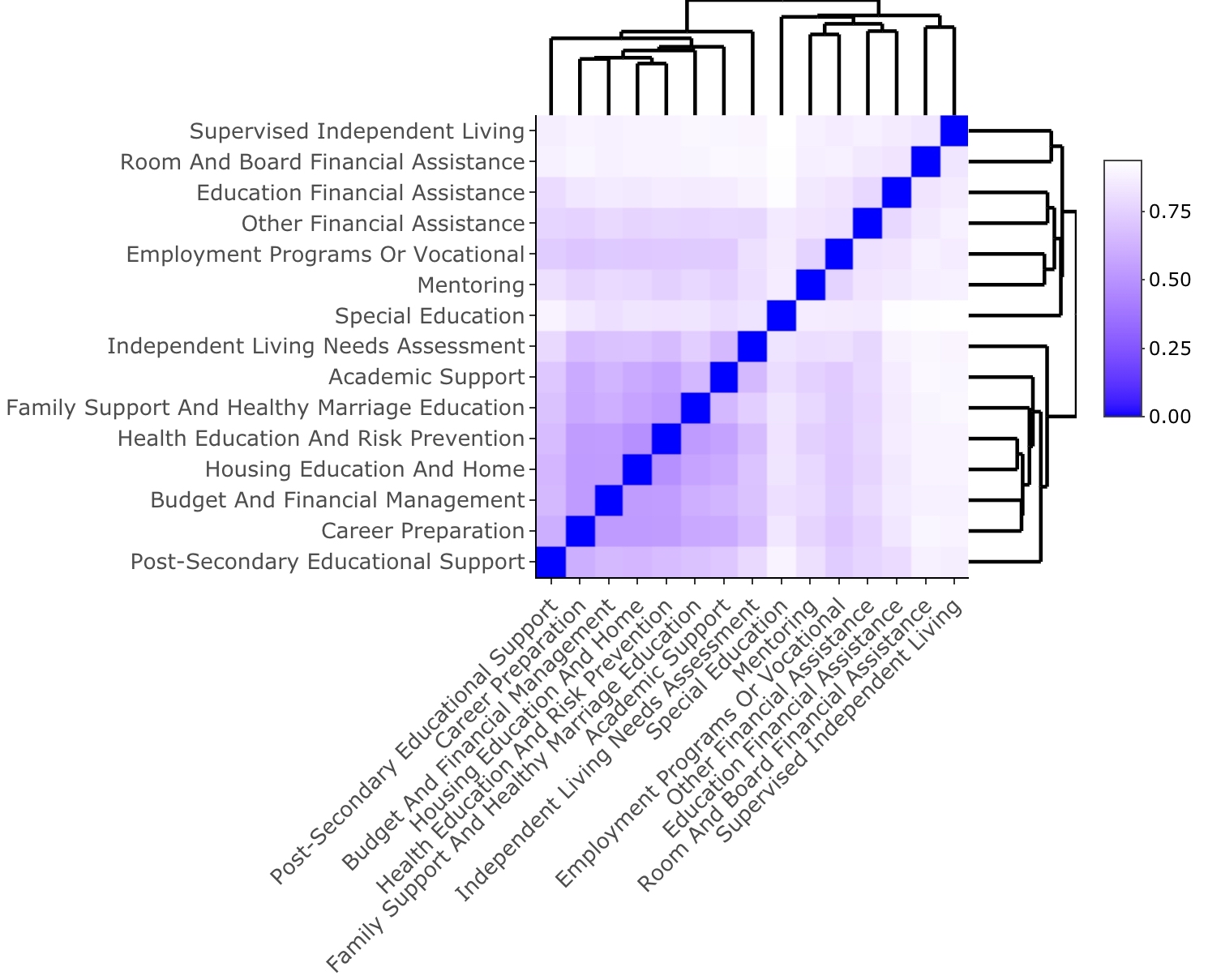}
    \caption{Jaccard similarity between each pair of Chafee services. Each cell represents the value of the Jaccard coefficient (described in the article body) measured between a pair of Chafee services. The higher Jaccard coefficient, the more blue the cell, and the higher the overlap in the youth who received both of those services. Services are also clustered using a hierarchical clustering algorithm, dendrograms are displayed that show these clustering patterns.}
    \label{fig:corr_services}
\end{figure}

Financial services also were more likely to be paired together with each other than with other services. More generally, we find clustering in the way services were allocated across youth.  Figure~\ref{fig:corr_services} presents the \emph{Jaccard coefficient} \citep{niwattanakul2013using} for each pair of Chafee services. The Jaccard coefficient is a ratio, for any pair of services, of the number of youth who received both services at least once, versus the number of youth who received at least one of those services. Mathematically, this is defined as the intersection of the set of youth who received each service over their union, i.e. $\frac{S_1 \cap S_2}{ S_1 \cup S_2}$, where $S_1$ and $S_2$ are sets of youth receiving two different services. If two services are given to the exact same set of youth, the Jaccard coefficient will be 1, in general, the higher the Jaccard coefficient, the stronger the overlap between the two services.

Figure~\ref{fig:corr_services} provides empirical validation of decisions made when operationalizing our dependent variable. First, it shows that Special Education is an outlier, in that it has limited overlap with any other services in the set of youth to whom it is allocated.  Second, Figure~\ref{fig:corr_services} shows that financial and Academic and Employment Support Services tend to be allocated to similar sets of youth, suggesting similar underlying factors that are potentially interesting to study in isolation are associated with their allocation. 

\begin{figure}[t]
    \centering
    \begin{tabular}{c c}
       \includegraphics[width=.45\textwidth]{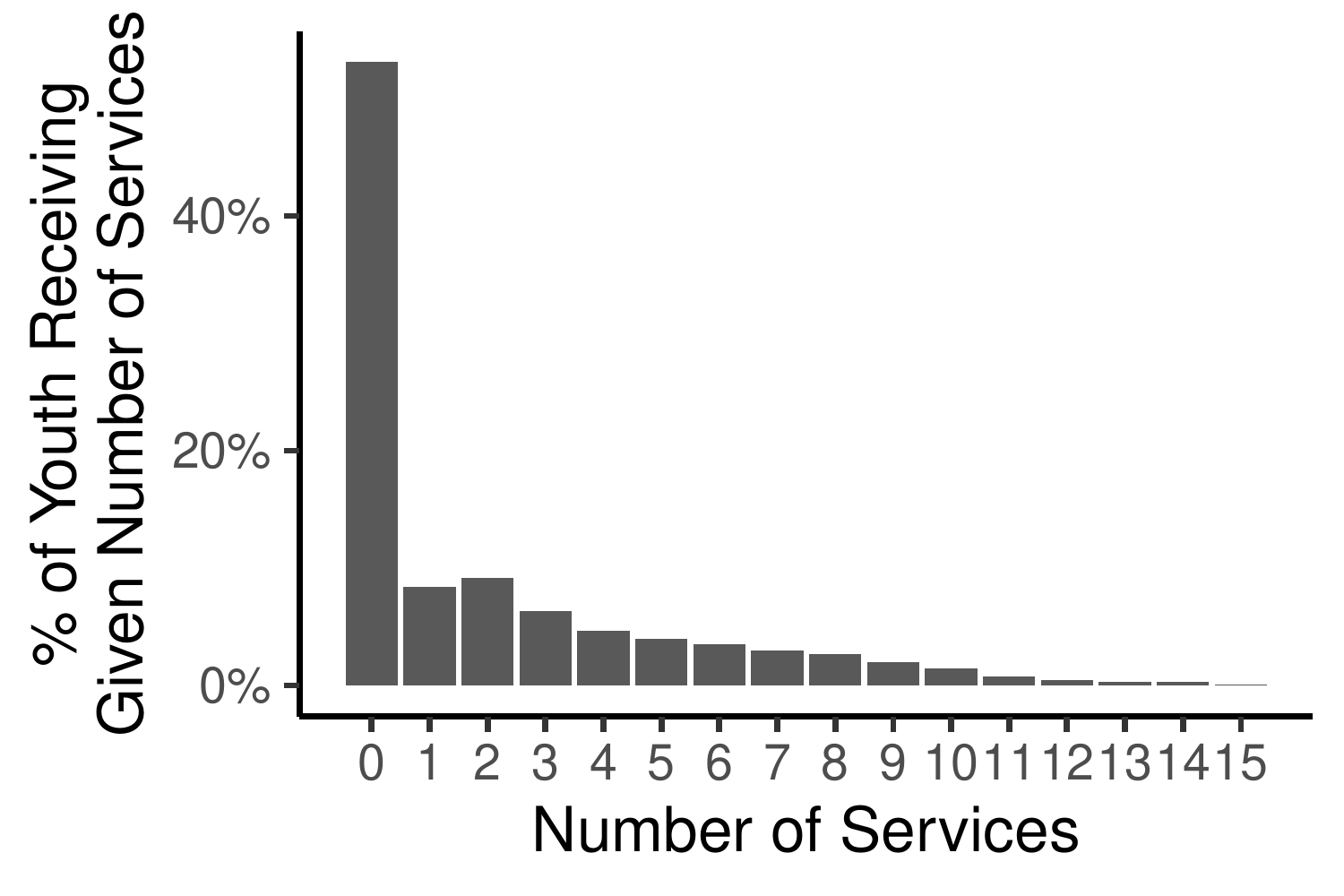}  & 
       \includegraphics[width=.45\textwidth]{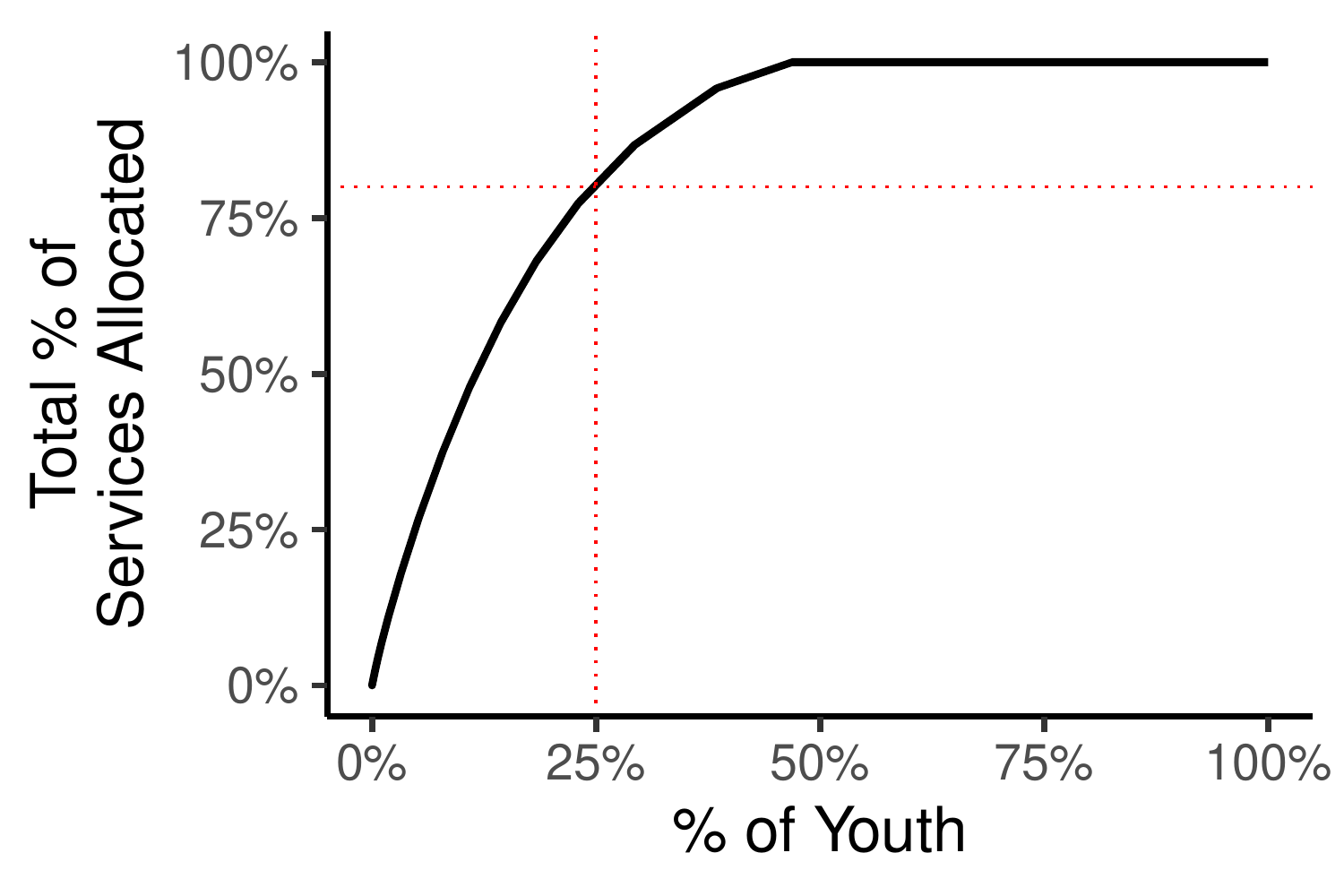}  \\
        (a) & (b)
    \end{tabular}
    \caption{A) The percentage of youth (y-axis) that received each possible number of unique services (x-axis).  B) The total percentage of all unique services (y-axis) allocated to a given percentage of youth (x-axis). Red dotted lines mark the fact that around 25\% of all youth receive around 80\% of all unique services allocated.}
    \label{fig:descriptives_1}
\end{figure}

Finally, it is important to draw a better understanding not only of the frequency of and correlations between distributions of services across all youth, but to also gain insight into patterns for individual youth. To this end, the distribution of services counts by child is shown in Figure~\ref{fig:descriptives)1}a). We find that 54\% of the youths received zero services, 82\% of youths received five services or less, and 93\% of youths received ten services or less. Only 5\% of youth receive more than 8 services. Figure~\ref{fig:descriptives_1}b) shows these disparities in allocation across youth from a different perspective, plotting cumulative services received against rank. Rank represents the top percentile of service users. For example, a rank of 0.2 would represent the top 20\% of youths with respect to number of services received. We can see that the top 25\% of youth received roughly 80\% of the services. 

These descriptive statistics broadly align with prior work and thus give us confidence that our results can speak to and extend these prior efforts. At the same time, the present a slightly simplified view of the results from \citet{chorYouthSubgroupsWho2018} and \citet{perezFactorsPredictingPatterns2020}. Specifically, the clustering they observed across which services tend to be allocated together seems to be apparent using more straightforward methods that just address correlations between pairs of services, rather than more complex latent class analysis methods. Similarly, while we, like prior work, identify a subset of high needs youth, we can state this in a simplified manner: we find that a small percentage of youth (25\%) receive the vast majority (80\%) of all services.

\subsection{Evaluation}

\begin{figure}[t]
    \centering
    \includegraphics[width=\textwidth]{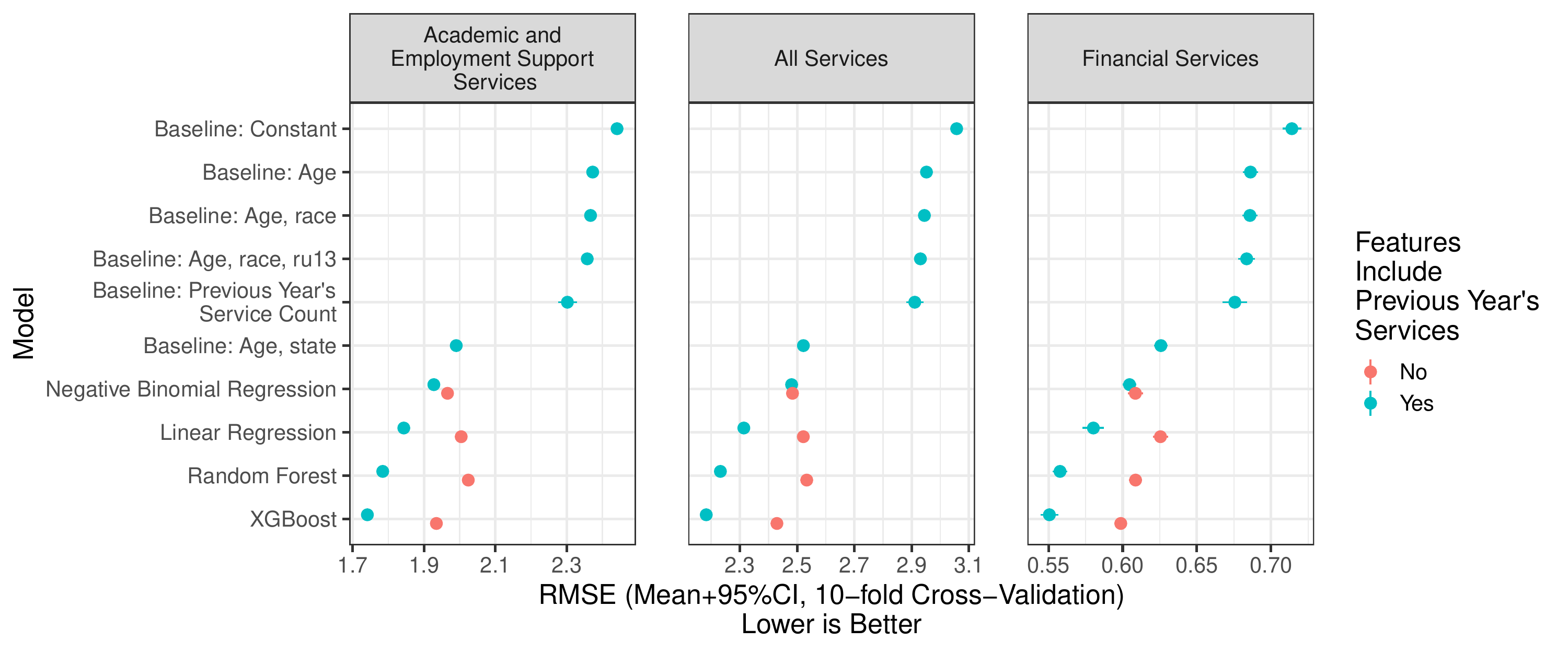}
    \caption{Results of our predictive experiment, using Root Mean Squared Error (RMSE; y-axis) as the outcome of interest. Each row represents a different prediction model, and each of the three sub-plots shows results for the three different dependent variables we analyzed, respectively.}
    \label{fig:predictions}
\end{figure}

We find, contrary to expectations from the work of \citet{salganikMeasuringPredictabilityLife2020}, that more complex models can significantly improve over plausible baseline methods in predictive the number of services received by youth.  More specifically, across all three dependent variables considered, the tree-based models we construct, using our full set of independent variables, significantly out-perform all other predictive models we considered.  On average, the GBT model was within 2.18 (95\% bootstrapps CI of [2.17,2.20]) services of the actual total number of unique services received by youth in 2018, and within 0.55 [.54,.56] and 1.74 [1.73,1.76] for Financial Services and Academic and Employment Support Services, respectively. As a point of comparison, this is an improvement of 25.5\%, 26.1\%, and 19.4\% over the descriptive baseline used by \citet{okpychReceiptIndependentLiving2015}, which took averages over age, race, and RU13. 

These results give us confidence that these more complex predictive models better capture important dimensions of variation in how services are allocated that may be useful in furthering our understanding of the data. However, this does \emph{not} mean that these models are of any potential use in \emph{making} decisions about how services are to be allocated. There are two reasons for this. First, as noted above, the prediction task we select uses several independent variables from 2018 to make predictions about the same time period; a model that sought to forecast how services should be allocated would need to do so without access to, for example, the youth's 2018 case goal. 

\begin{figure}[t]
    \centering
    \includegraphics[width=\textwidth]{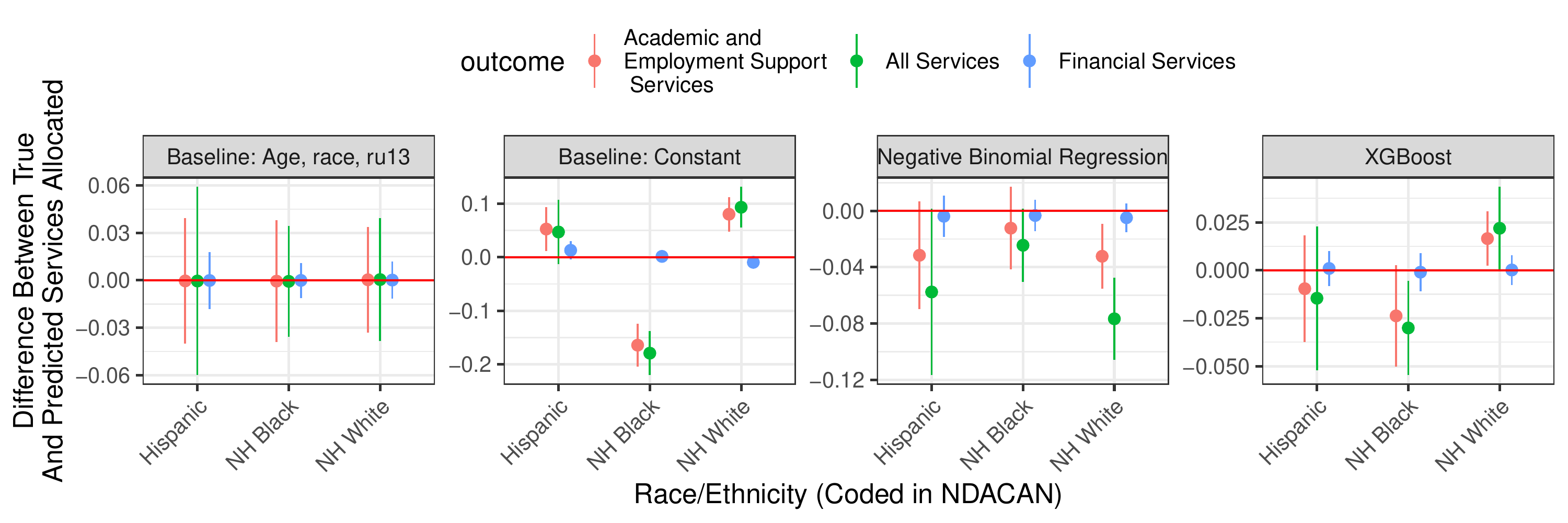}
    \caption{Results of our predictive experiment, using Mean Error (y-axis) as the outcome of interest, with results differentiated by race/ethnicity (y-axis; only a subset of race/ethnicities coded in NDACAN are used here). Each separate subplot represents a different model (only a subset of models are plotted here) and colors represent the different independent variables considered}
    \label{fig:race_diff}
\end{figure}

Second, Figure~\ref{fig:race_diff} shows that the most accurate model, as measured by RMSE, is not the most equitable model. Specifically, the GBT model that was most predictive also would, if used as an allocation mechanism, provide significantly more services to non-Hispanic White youth than it would to non-Hispanic Black youth, or to Hispanic youth, for two of the three independent variables considered. In contrast, the simple descriptive model developed by \cite{okpychReceiptIndependentLiving2015}, as well as the non-tree based negative binomial model we assessed here, would show no such favoritism.  This observation underscores the importance of using forensic social science as a tool to help guide theoretical development and additional empirical work, rather than as a mechanism for decision-making in and of itself.

\subsection{Model Exploration}

\begin{figure}[t]
    \centering
    \includegraphics[width=\textwidth]{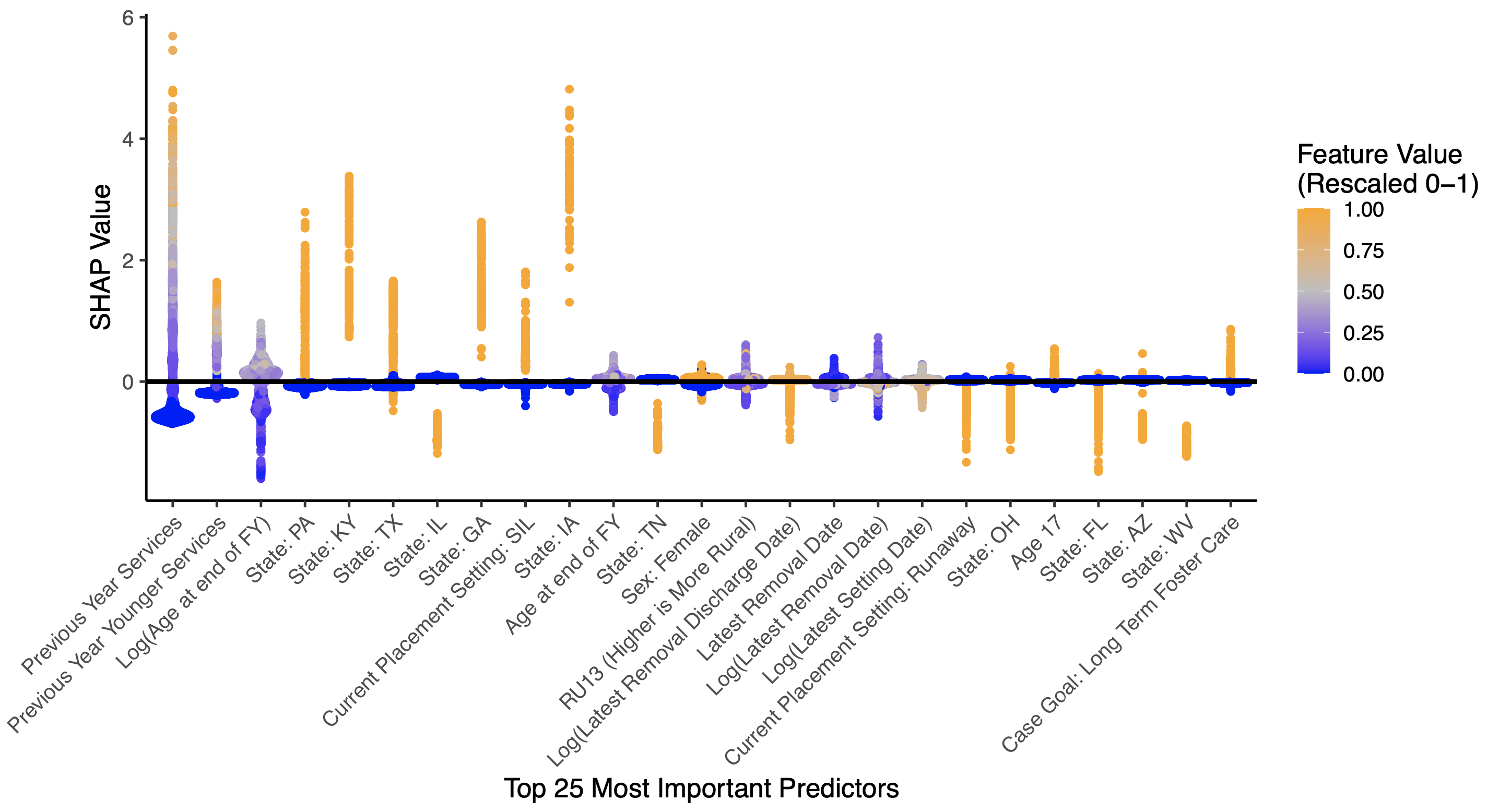}
    \caption{SHAP values (y-axis) for 5000 randomly selected points (each dot)for the 25 most important independent variables in terms of mean absolute SHAP values (x-axis). Variables are ordered by their average absolute impact, from highest (left) to lowest (right). Color represents the magnitude of the feature value; all features are rescaled from 0 (blue, lowest value) to 1 (orange,highest value).}
    \label{fig:shap_all}
\end{figure}

We select the best-performing GBT model to explore for insights into factors correlated with service allocation. In this section, we explore only factors associated with predictions for the \emph{All Services} outcome, as we find similar results for the other two outcome variables. Figure~\ref{fig:shap_all} shows that the most important variables for this model and suggests that these variables primarily fall into one of four categories: the number of services a youth received in the previous year, the youth's age, the youth's length of time in care, and the state in which the youth resides.  

Figure~\ref{fig:shap_all} plots the top 25 independent variables as determined by their mean absolute SHAP value - that is, the 25 variables that had the strongest average effect on the outcome variable.  The plot shows, for 5000 randomly sampled youth on each independent variable, the SHAP value for that youth on the y-axis, and displays using color the value of the feature itself. So, for example, the first element on the x-axis shows that youth with more services in 2017 (as evidenced by the orange color of the points) received far more services (higher values on the y-axis).

It is unsurprising, given how Chafee services are allocated, that age and services received in the prior year would impact the number of services youth received. Similarly, outside of the four primary categories of features we note above, prior work has shown that sex and urbanicity are important factors \citep{okpychReceiptIndependentLiving2015,perezFactorsPredictingPatterns2020,chorYouthSubgroupsWho2018}. Finally, placement settings and case goals that appear in Figure~\ref{fig:shap_all} have natural explanations; for example, a youth placed in supervised independent living (SIL) is of course more likely to have received this service.

\begin{figure}[t]
    \centering
    \includegraphics[width=\textwidth]{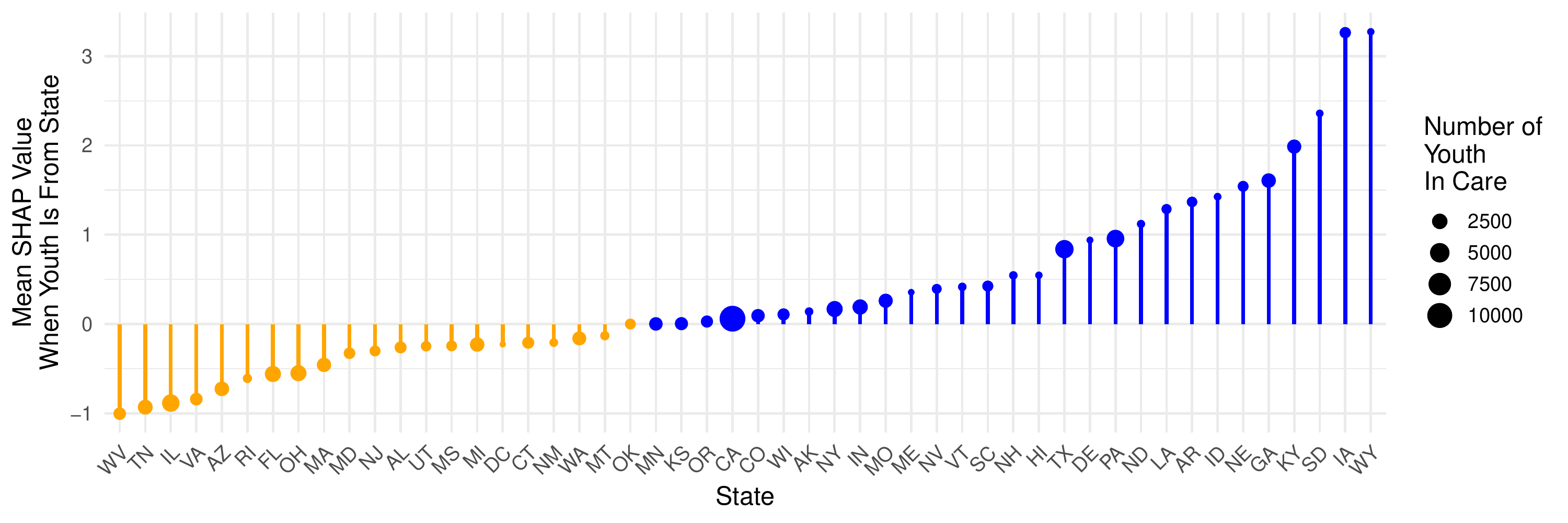}
    \caption{The mean SHAP value (y-axis) for youth residing in a given state (x-axis). Point size represents the number of observed youth.}
    \label{fig:shap_state}
\end{figure}

\begin{figure}[t]
    \centering
    \includegraphics[width=\textwidth]{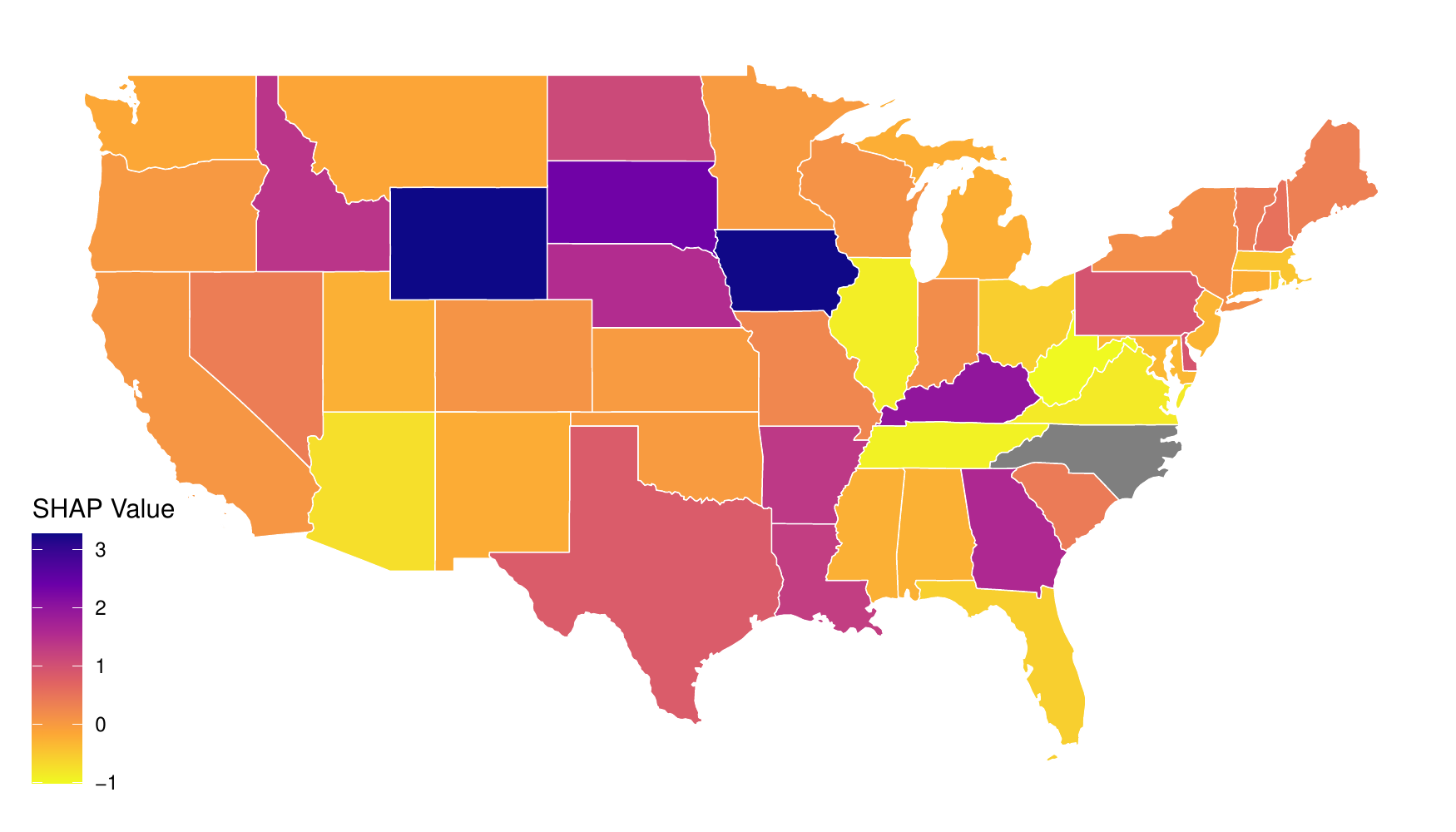}
    \caption{The mean SHAP value (color) for youth residing in a given state.}
    \label{fig:shap_state_loc}
\end{figure}

There are, however, two classes of variables that have not seen significant attention in prior work that stand out in our model exploration. First, while prior work has acknowledged the importance of state-level variation, little has been done to explore this variation or why it may exist. Our findings shed some light on how service allocation varies across states.

Figure~\ref{fig:shap_state} shows the mean SHAP value for each state, and Figure \ref{fig:shap_state_loc} serves as a spatial visualization of potential trends across some groupings of neighboring states based on these values. In particular, there are five groupings of two or more states with similar SHAP values:
\begin{enumerate}
    \item VA, WV, and TN, where youth received fewer services on average;
    \item CA and OR, neither of which had much of an effect on the number of services youth received;
    \item ME, VT, and NH, all of which have SHAP values indicating a slight increase in the number of services youth received;
    \item LA and AR, all of which have SHAP values that point to a moderate increase in the number of services youth received on average; and
    \item WY, SD, and IA, the three states with the highest SHAP values.
\end{enumerate}

It is unclear why some neighboring states might have a similar influence on service allocation, though theories regarding inter-jurisdictional policy diffusion and budget spillover point to some areas for future investigation. First, there are many reasons that government spending in one state might influence that in another, such as concerns about appearing too generous or austere relative to other states or constituents’ tendencies to compare state politicians to those in neighboring states \citep{baickerSpilloverEffectsState2005}. In the case of medical spending, \citet{baickerSpilloverEffectsState2005} identified interstate migration as having the most considerable influence on budget spillover between states. They stressed the importance of evaluating a range of other neighborliness metrics, including geographic proximity, similarities in demographic composition, per capita income, and population size. Exploring these variables might help identify possible causes for our finding of clustering between contiguous states.

Second, prior research has revealed interdependence between neighboring states whereby one state’s economic growth is dependent on productive government spending in another; conversely, budget cuts in one state can have detrimental effects on others within proximity \citep{ojedeDirectIndirectSpillover2018}. Youths’ needs for room and board, other financial assistance, and education and training vouchers might be influenced by broader economic trends related to the housing market, labor market, and public higher education system, all aspects of a state’s economy that are likely dependent on the economies of nearby states. For instance, \citeauthor{ojedeDirectIndirectSpillover2018}’s (2018) findings demonstrated how public spending on higher education could contribute to growth in per capita income within neighboring states. Equally important are common economic factors, such as regional economic downturns, that may impact spending across several contiguous states in a similar manner \citep{baickerSpilloverEffectsState2005}.

Policy diffusion is yet another area worthy of exploration. In the past, states have exhibited regional clustering patterns related to the length of time it took them to adopt child welfare policies according to federal mandates \citep{lloydsiegerVariationStatesImplementation2020}. Although we identified what looks like regional clustering between neighboring states, it is crucial to recognize that contiguity is not the only variable that plays a role in the diffusion of policies and that states outside of these clusters with similar SHAP values (e.g., NV and VT) should also be evaluated for potential similarities in policies and regulations \citep{karchEmergingIssuesFuture2007}. For example, policymakers in one state may imitate another state’s policy initiatives based on shared political commitments or similarities in demographic composition \citep{karchEmergingIssuesFuture2007}, or philanthropic foundations may endorse policy diffusion across states and play a role in determining how policies are implemented once passed \citep{bushouseIntermediaryRolesFoundations2018}. In these cases, the spread of policies may have more to do with the networks established by politicians and interest groups than with geographic proximity \citep{bushouseIntermediaryRolesFoundations2018}. Finally, it is worth examining the potential influence of internal or external events that differentially shaped state policy environments over the past couple of decades in ways that may have had long-lasting effects on how services were allocated in the following years \citep{jenningsMoralPanicsPunctuated2020}. 

\begin{figure}[t]
    \centering
    \includegraphics[width=.7\textwidth]{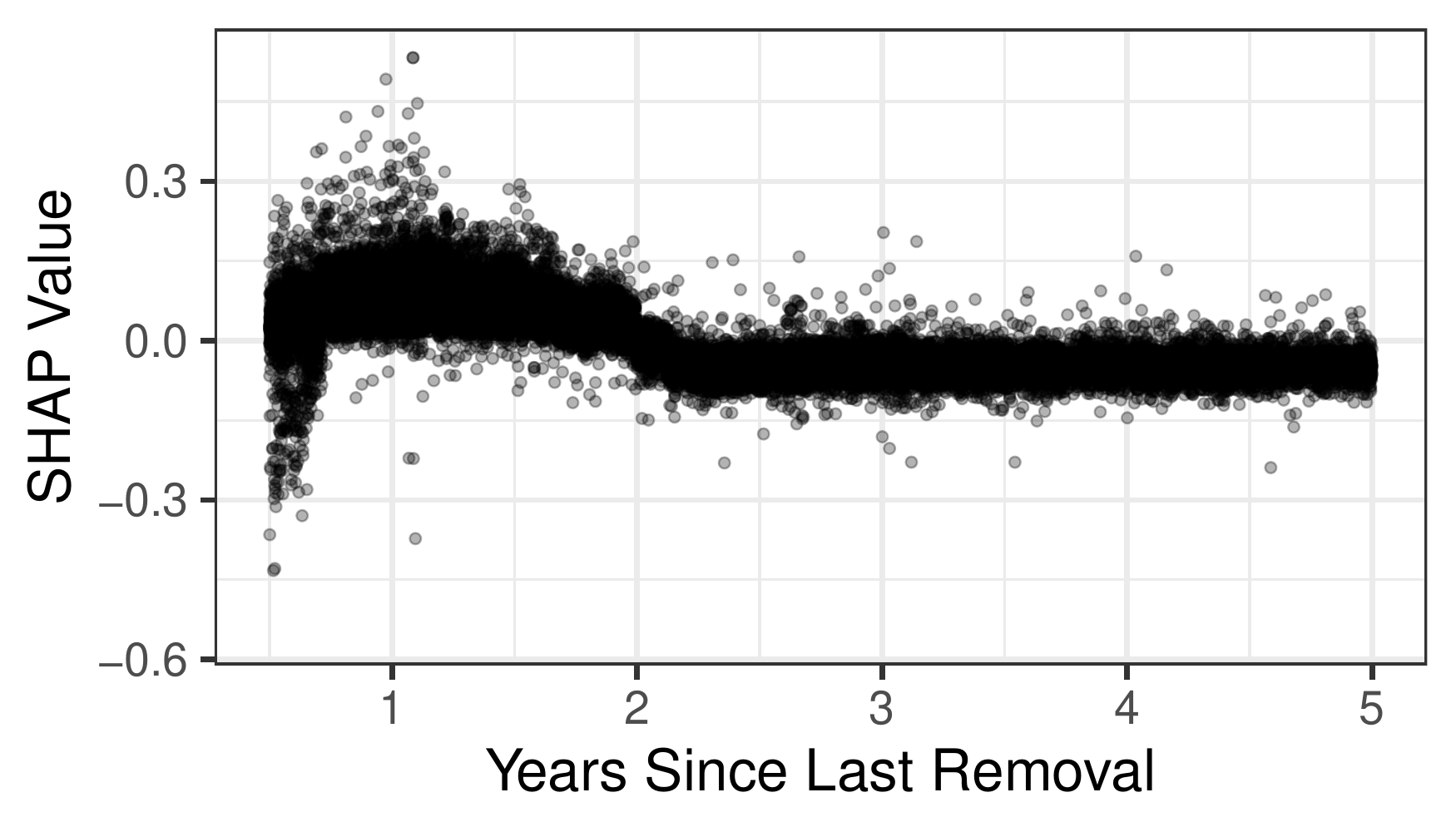}
    \caption{Each point represents a single youth in the dataset and the SHAP value (y-axis) for the independent variable representing the number of days since they were last removed from their home plotted against the actual value for this variable (x-axis).}
    \label{fig:shap_lenstay}
\end{figure}

In addition to state, no prior work has yet explored the relationship between length of latest stay in care and services received. Figure~\ref{fig:shap_lenstay} shows that there is a non-linear relationship between the expected number of services a youth receives and how long they have been in care. One might presume that youth with longer stays in foster care have more time to benefit from the range of independent living services available to foster youth. Past research has posited that eligible youth who are in foster care extend their stays to leverage more services, which may result in a correlation between extended foster care and the number of services \cite{courtney2014findings}, Notably, in the Midwestern Study \cite{courtney2011receipt}, transition age youth report that the quantity of services is consistent with length of stay.

\section{Discussion and Conclusion}
Presently, there is robust debate in the child welfare literature about whether computer science methods can help understand the risks and needs of youth involved in the child welfare system. One major concern is the fear that algorithmic decision-making may appear more neutral, but amplify bias that is embedded in data. We demonstrate that algorithms that maximize for fit may indeed decrease equity in service allocation problems. 

Social sciences often gives preference to theory-driven models and computer science often prefers data-driven atheoretical models. However, in light of the impact of practice assumptions when using human services data, forensic science may provide an approach that balances these lenses. 
We demonstrate that the iterative approach to clustering services that are theoretically linked, paired with a data science approach that helps uncover the characteristics in each cluster, informs new research explanations and questions. This might help in the development of theoretical perspectives. For instance, what common factors exist between states that are similar in service allocation, and what can that tell us about the role of policy?

A forensic approach may also help explore data for best-practice implications. For instance, relationships between multiple variables such as state, age, and service types, and the impact of these on an outcome such as homelessness, may help point us to models of policy-practice success that deserve further exploration.  A practice theory lens is important for knowing where to start and the possible causes of interactions, but the ability to explore the data allows for space to develop new hypotheses from the data.

Finally, while much of the current conversations about advanced data analytics centers on decision-making ethics using prediction, this type of analysis also offers opportunities to increase transparency and inform decision-making about macro impacts. For instance, state as a predictor of service allocation raises questions about the impact of statewide policy and practice, and points to questions about funding levels, political values, and geographic service distribution, contrary to the common assumption that youth needs and demographics drive service delivery.  Similarly, the associations between youth age and service delivery raise questions about whether there exists ideal age-based timing for services and how that timing influences outcomes.  In short, computational social science strategies may help us determine which states or policies are most in need of service, versus which youth are most in need of services, which provides a new way of thinking about predictive analytics for child welfare.





\bibliography{bib}
\bibliographystyle{apalike}


\end{document}